\documentclass[%
aip,
rsi,%
amsmath,
amssymb,
reprint,%
floatfix
]{revtex4-1}

\usepackage{graphicx}
\usepackage{dcolumn}
\usepackage{bm}

\usepackage{xcolor}

\begin{document}

\preprint{AIP/123-QED}

\title[Nano-oscillator-based classification with a machine learning-compatible architecture]{Nano-oscillator-based classification with a machine learning-compatible architecture}

\author{D. Vodenicarevic}
\affiliation{ 
Centre for Nanoscience and Nanotechnology, CNRS, Univ Paris-Sud, Universit\'e Paris-Saclay,\\
rue Andr\'e Amp\`ere, 91405 Orsay, France
}%
\author{N. Locatelli}%
\affiliation{ 
Centre for Nanoscience and Nanotechnology, CNRS, Univ Paris-Sud, Universit\'e Paris-Saclay,\\
rue Andr\'e Amp\`ere, 91405 Orsay, France
}%
\author{J. Grollier}%
\affiliation{ 
UMP CNRS/Thales, Univ Paris-Sud, Universit\'e Paris-Saclay,  \\
1 Avenue Augustin Fresnel, 91767 Palaiseau, France
}%
\author{D. Querlioz}%
\affiliation{ 
Centre for Nanoscience and Nanotechnology, CNRS, Univ Paris-Sud, Universit\'e Paris-Saclay,\\
rue Andr\'e Amp\`ere, 91405 Orsay, France
}%

\date{\today}

\begin{abstract}
Pattern classification architectures leveraging the physics of coupled nano-oscillators have been demonstrated as promising alternative computing approaches, but lack effective learning algorithms. In this work, we propose a nano-oscillator based classification architecture where the natural frequencies of the oscillators are learned linear combinations of the inputs, and define an offline learning algorithm based on gradient back-propagation. Our results show significant classification improvements over a related approach with online learning. We also compare our architecture with a standard neural network on a simple machine learning case, which suggests that our approach is economical in terms of numbers of adjustable parameters.  The introduced architecture is also compatible with existing nano-technologies: the architecture does not require changes in the coupling between nano-oscillators, and it is tolerant to oscillator phase noise.
\end{abstract}

\maketitle

\section{Introduction}

Heavy demand for cognitive processing of large amounts of data using machine learning software is increasing the strain on data center energy consumption and hand-held device battery autonomy ~\cite{mills_cloud_2013}. 
Many alternative computing architectures are therefore being developed to match the specificities of such tasks by trading precision and sequential computation speed for increased parallelism and improved energy efficiency~\cite{schneider_deeper_2017}. 
Most of these approaches rely on digital operations to approximate the non-linearities at the heart of cognitive algorithms.
On the contrary, the human brain successfully performs cognitive tasks in an energy-efficient fashion using slow, noisy, variable, unreliable but naturally non-linear neurons
~\cite{hasler_special_2017}.  
It is therefore highly attractive to mimic this approach of the brain, and to leverage non-linear device physics to perform cognitive tasks more efficiently~\cite{schuman_survey_2017,torrejon_neuromorphic_2017}.
This idea takes special sense with the recent advances in nanotechnology, which provide fast, compact and integrable nano-oscillators, with coupling and synchronization capabilities: nano-oscillators based on mechanical vibrations~\cite{hoppensteadt_synchronization_2001}, oxide phase changes~\cite{sharma_phase_2015}, Josephson junctions~\cite{ovchinnikov_networks_2013}, or spintronics~\cite{csaba_computational_2013,locatelli_spintronic_2015}.
Multiple recent works draw inspiration from the oscillatory activity observed in the brain at different scales~\cite{wang_neurophysiological_2010} to naturally perform pattern classification by leveraging the rich dynamics of networks of coupled oscillators ~\cite{vassilieva_learning_2011,levitan_associative_2013,nikonov_coupled-oscillator_2015,cotter_computational_2014,parihar_computing_2016,romera_training_2017,vodenicarevic_nanotechnology-ready_2017,Ignatove1700849,Fange1601114,csaba2018perspectives}. 

However, this approach has to overcome a difficult challenge. The natural frequencies of nano-oscillators can usually be tuned, by applying current or voltage biases, but dynamically adjusting inter-oscillator coupling strengths requires heavy circuitry \cite{nikonov_coupled-oscillator_2015}. As a consequence, traditional learning methods used for training neural networks, which usually consist in adjusting the couplings between non-linear units, are not very adapted to oscillator-based classifiers.

An attractive alternative approach is the oscillator-based classifier initially proposed in a mathematical context by Vassilieva~\textit{et~al.}~\cite{vassilieva_learning_2011}, and adapted to nano-oscillator technologies in~\cite{vodenicarevic_nanotechnology-ready_2017}.
The online learning procedure of this network  involves reading the synchronization state of various pairs of oscillators, and adjusting their natural frequencies in order to reinforce expected synchronizations, and weaken unexpected ones for each known input example presented to the oscillator network. 
This approach was recently validated  experimentally by Romera~\textit{et~al.}, who  achieved spoken vowel recognition using a network of four coupled spin-torque nano-oscillators~\cite{romera_training_2017}.
 
In this work, we propose an ``extended'' nano-oscillator-based classifier architecture that retains the nanotechnology-compatibility of the reference approach of~\cite{vassilieva_learning_2011}, but 
is trained using gradient descent, the standard algorithm of machine learning. This allows our architecture to capitalize on developments realized for more conventional forms of neural networks.
We show that our extended classifier achieves better classification results than the reference classifier on a simple task. We then run a standard classification benchmark of the extended classifier against a classical neural network and investigate the robustness to phase noise and the scalability of the extended architecture. This allows us to discuss the advantages and drawbacks of our approach.

\section{Definition of the oscillator-based classification architectures}

\begin{figure}[ht!]
\includegraphics[]{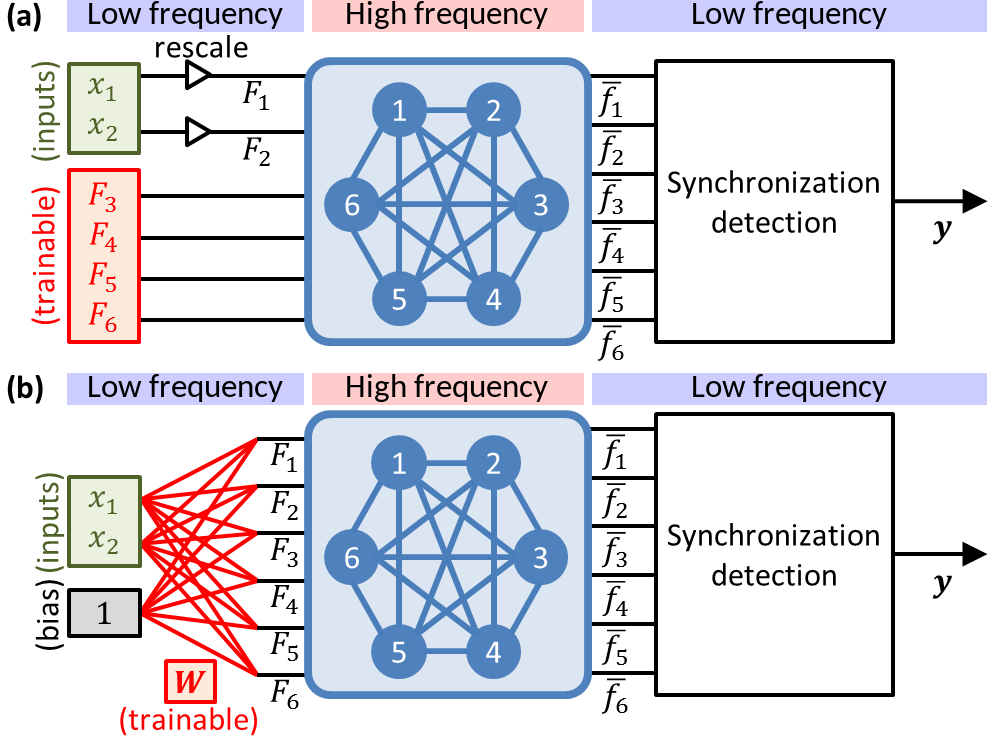}
\caption{(a) Schematic of the reference classifier with 2 input oscillators (1 and 2) and 4 tuning oscillators (3,4,5 and 6). (b) Schematic of the extended classifier with 6 oscillators.} 
\label{fig:archis}
\end{figure}

Fig.~\ref{fig:archis}(a) shows a pattern classification architecture, 
subsequently called reference classifier,
similar to the one originally proposed by Vassilieva~\textit{et~al.}~\cite{vassilieva_learning_2011}. 
This classifier consists of a set of $N$ oscillators coupled all-to-all through constant uniform weak couplings. Among the $N$ oscillators, $N_I$ act as input oscillators (oscillators 1 and 2 in Fig.~\ref{fig:archis}(a)), and the $N_T$ others act as tuning nano-oscillators (oscillators 3, 4, 5 and 6 in Fig.~\ref{fig:archis}(a)).
The input vector $\bm{x}$ to be classified is presented by setting the natural frequencies of the input oscillators ($F_1,F_2$ in Fig.~\ref{fig:archis}(a)) accordingly. The input oscillators induce synchronizations in the oscillator network, which can be deduced from the average observed frequencies $\bm{\bar f}$ of the oscillators~\cite{vodenicarevic_synchronization_2016,romera_training_2017}. The resulting list of synchronized pairs of oscillators then yields a binary output vector $\bm{y}$ which represents the class that the system has associated to the presented pattern. 
The response of the classifier can be trained to perform a given classification task by adjusting the natural frequencies of the tuning oscillators ($F_3,F_4,F_5,F_6$ in Fig.~\ref{fig:archis}(a)).

This reference architecture provides many benefits for implementation with nanotechnology: resilience to device variability, to noise, to device non-linearity, and it has good scaling properties \cite{vodenicarevic_nanotechnology-ready_2017}. {\color{black}Moreover, high frequency signals are present only within the oscillator network, while natural frequency control inputs, and synchronization detection outputs are low frequency signals, which eases implementation.}

The limitation of the architecture, by contrast, is that it only provides $N_T$ tunable parameters, which limits its classification capabilities for low numbers of oscillators. Moreover, the classifier response is restricted by strong symmetries arising from the fact that all oscillators are identical and only defined by their natural frequencies. 

Overcoming such limitations requires rethinking the classification architecture. 
{\color{black}
In Fig.~\ref{fig:archis}(b), we 
introduce
an extended version of the classifier that uses a weight matrix formalism, compatible with existing neural network frameworks. 
The natural frequencies of all oscillators are now trainable linear combinations of the inputs $\bm{x}$ and a constant bias according to: $\bm{F}=\bm{W} \begin{bmatrix}\bm{x} \\ 1 \end{bmatrix}  $, where $\bm{W}$ is the trainable weights matrix. 
This extended classifier therefore erases the difference between input and tuning oscillators as all the $N$ oscillators in the network play a similar role. 
}
As each natural frequency is a potentially different linear combination of the inputs, the symmetries present in the reference approach of \cite{vassilieva_learning_2011} are broken and $\left(\dim\left(\bm{x}\right)+1\right) \times N$ tuning parameters are made available. 
As the high-frequency couplings between oscillators are kept uniform and constant, the extended classifier remains compatible with the constraints of nanotechnology, only adding a low-frequency linear combination step to the inputs, which has already been implemented in CMOS~\cite{furber_overview_2013,merolla_million_2014} or with programmable resistive devices~\cite{lin_physical_2016}. 

\section{Presentation of the learning algorithms}

The iterative online learning algorithm defined by Vassilieva~\textit{et~al.} \cite{vassilieva_learning_2011} and later adapted to experiments by Romera~\textit{et~al.} \cite{romera_training_2017} consists in presenting a set of training input examples $\{\bm{x}^{(m)}\}_{m=1..M}$ with known expected classification binary vectors $\{\bm{\hat{y}}^{(m)}\}_{m=1..M}$, and adjusting the natural frequencies according to the measured synchronizations. For each presented example $(m)$, the resulting list of synchronized pairs of oscillators is compared to the expected one. 
{\color{black}
The learning algorithm then aims at weakening the synchronizations of pairs of oscillators that are not expected for the example $(m)$, by slightly pushing their natural frequencies apart, and at promoting desynchronized pairs that are expected to be synchronized, by slightly pulling their natural frequencies closer together. 
This behavior is achieved by applying linear updates that adjust the natural frequencies of the output oscillators according to the sum of observed frequency differences, weighted by a factor depending if the oscillators are to or not to synchronize\cite{vassilieva_learning_2011}.
}
However, this algorithm only acts on the natural frequencies of oscillator pairs used as outputs and compared to the expected example outputs, which does not allow having extra non-output oscillators to increase the computational power of the system at the same output dimensionality. More precisely, the online learning algorithm restricts the number of trained oscillators to a maximum of $2\times \dim(\bm{y})$. Moreover, the online learning algorithm is 
``greedy''
as it acts on each pair of oscillators assuming they are isolated and do not influence the other oscillators of the network. As all oscillators interact together through complex network dynamics, this assumption can lead the algorithm to suboptimal results.

\begin{figure}[ht!]
\includegraphics[]{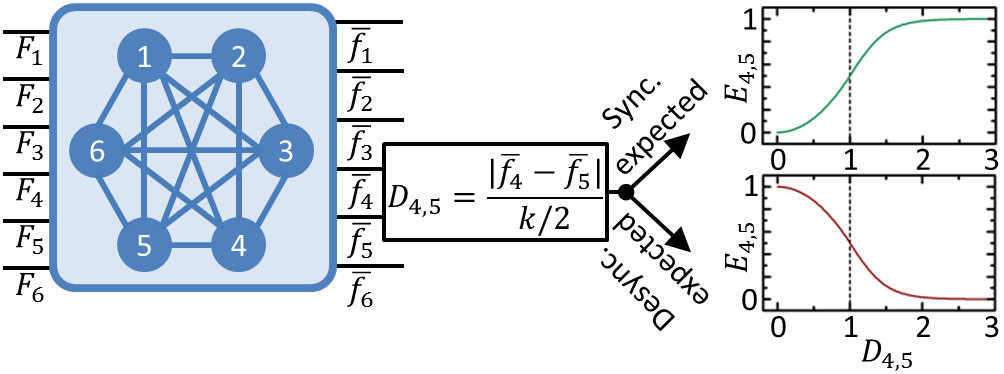}
\caption{Schematic of the  differentiable error function depending on the expected synchronization states for a given input example in the case of a binary classification task with output pair (4,5).}
\label{fig:err}
\end{figure}

In this work, we introduce an offline learning algorithm inspired by standard machine learning techniques that overcomes the limitations of the online learning algorithm by iteratively minimizing, through gradient descent, the total error function:
\begin{equation}
E^{\mathrm{tot}}= \frac{1}{M}\sum_{m=1}^M \sum_{\substack{\text{output} \\ \text{pairs}~(i,j)}} E_{i,j}^{(m)}
\end{equation}
where $E_{i,j}^{(m)}$ is the error on the output pair $(i,j)$ for the training example $(m)$. Fig.~\ref{fig:err} illustrates the value of $E_{i,j}^{(m)}$, which depends on the expected synchronization state of the pair $(i,j)$ for the example $(m)$ and a measure of the actual desynchronization $D_{i,j}^{(m)}$ of the pair $(i,j)$ after the example $(m)$ is presented, defined by:
\begin{equation}
D_{i,j}^{(m)}=\frac{|\bar f_i^{(m)} - \bar f_j^{(m)}|}{k/2} \,,
\end{equation}
{\color{black} where $k$ is the uniform coupling strength between the oscillators.}
$(i,j)$ are considered synchronized if $D_{i,j}^{(m)}\leq 1$ and desynchronized otherwise.
If the oscillators $(i,j)$ are expected to be synchronized for the presented example $(m)$, we define:
\begin{equation}\label{eq:errsync}
E_{i,j}^{(m)} = \begin{cases}
\displaystyle{ \frac{1}{2} {\left( D_{i,j}^{(m)} \right)}^2 \quad \mathrm{if} \quad D_{i,j}^{(m)}\leq 1 }\\[1.2em]
\displaystyle{ \frac{1}{1+e^{-4\left(D_{i,j}^{(m)}-1\right)}} \quad \mathrm{if} \quad D_{i,j}^{(m)} > 1 }
\end{cases} \,,
\end{equation}
or if the synchronization $(i,j)$ is not expected for the presented example $(m)$:
\begin{equation}\label{eq:errdesync}
E_{i,j}^{(m)} = \begin{cases}
\displaystyle{ 1-\frac{1}{2} {\left( D_{i,j}^{(m)} \right)}^2 \quad \mathrm{if} \quad D_{i,j}^{(m)}\leq 1 }\\[1.2em]
\displaystyle{ 1-\frac{1}{1+e^{-4\left(D_{i,j}^{(m)}-1\right)}} \quad \mathrm{if} \quad D_{i,j}^{(m)} > 1 }
\end{cases} \,.
\end{equation}

{\color{black}
Computing the total error gradient with respect to the natural frequencies $\frac{\partial E^\mathrm{tot}}{\partial \bm{F}}$ for the reference classifier or with respect to the weight matrix elements $\frac{\partial E^\mathrm{tot}}{\partial \bm{W}}$ for the extended classifier requires in both cases the derivatives $\frac{\partial E_{i,j}^{(m)}}{\partial \bm{F}}$, which can be expanded using the chain rule:
\begin{equation}
\frac{\partial E_{i,j}^{(m)}}{\partial {F_a}}=
\frac{\partial E_{i,j}^{(m)}}{\partial {\bar f^{(m)}_i}}
\frac{\partial {\bar f^{(m)}_i}}{\partial F_a}
+
\frac{\partial E_{i,j}^{(m)}}{\partial {\bar f^{(m)}_j}}
\frac{\partial {\bar f^{(m)}_j}}{\partial F_a} \,,
\end{equation}
where the derivatives $\frac{\partial E_{i,j}^{(m)}}{\partial \bm{\bar f^{(m)}}}$ are obtained by differentiating equations~(\ref{eq:errsync},\ref{eq:errdesync}).
}

Continuing the gradient computation requires the derivative of the average frequencies with respect to the natural frequencies of the oscillators: $\frac{\partial \bm{\bar f}}{\partial \bm{F}}$. The partial derivatives $\frac{\partial \bm{\bar f}}{\partial \bm{F}}$ correspond to the Jacobian matrix of the oscillator network operator depicted as a blue rounded square in Figs.~\ref{fig:archis} and \ref{fig:err}, and translate the rich dynamics of all the oscillators interacting together. 

Throughout the paper, we model the oscillator network using the Kuramoto model, with which many nanooscillators can be described \cite{vodenicarevic_nanotechnology-ready_2017}. 
In order to compute the Jacobian offline, we apply a simple Euler numerical integration scheme following the update equations for the oscillator instantaneous frequencies $\bm{f}(t)$ and phases $\bm{\theta}(t)$ with a time step $\mathrm{dt}=10^{-11}\mathrm{s}$, and total simulation time $T=1\mathrm{\mu s}$:
\begin{equation}\label{eq:euler}
\begin{cases}
\displaystyle{f_i(t)= F_i + k\sum_{i=1}^N{ \sin{\left(\theta_j(t) - \theta_i(t)\right)} }}\\[1.2em]
\displaystyle{\theta_i(t+\mathrm{dt})= \theta_i(t) + 2\pi \mathrm{dt} f_i(t) }\\[0.5em]
\end{cases} \,.
\end{equation}

\begin{figure}[ht!]
\includegraphics[]{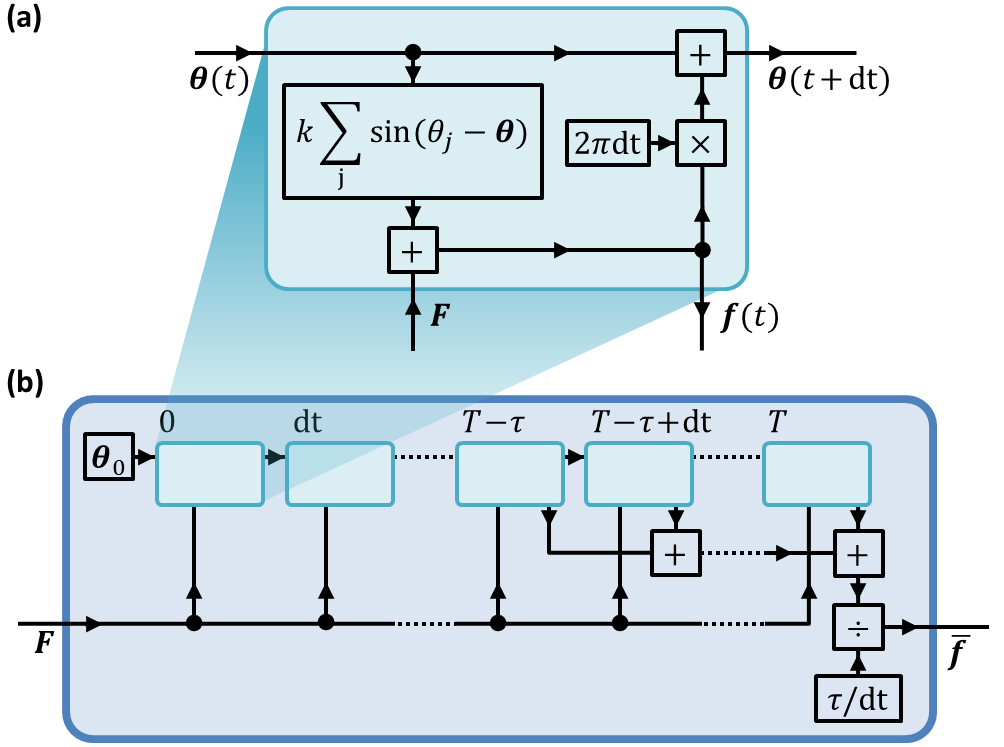}
\caption{(a) Schematic representation of the differentiable recurrent cell reinterpretation of an Euler integration step of the Kuramoto model. (b) A fully differentiable oscillator network operator taking a vector of natural frequencies $\bm{F}$ as input, and outputting a vector of mean frequencies $\bm{\bar f}$.}
\label{fig:backprop}
\end{figure}

As the state of the system at time $t+\mathrm{dt}$ is a continuously differentiable function of the state at time $t$, we translate the numerical update equations~(\ref{eq:euler}) into a continuously differentiable operator as shown in Fig.~\ref{fig:backprop}(a). This update operator is then stacked in time as shown in Fig.~\ref{fig:backprop}(b) and the instantaneous frequencies $\bm{f}(t)$ are averaged on the last $\tau=0.5\mathrm{\mu s}$ to represent a continuously differentiable expression of the oscillator network operator taking $\bm{F}$ as input and outputting $\bm{\bar f}$.
{\color{black}
The derivatives are updated at each simulation step following:
\begin{equation}
\begin{cases}
\displaystyle{
\frac{\partial f_i(t)}{\partial F_a} =
\delta_{i,a} + k\sum_j{ \left( 
\frac{\partial \theta_j(t)}{\partial F_a}
-
\frac{\partial \theta_i(t)}{\partial F_a}
\right) \cos\left(\theta_j(t)-\theta_i(t) \right) } }\\
\displaystyle{\frac{\partial \theta_i(t+dt)}{\partial F_a}=
\frac{\partial \theta_i(t)}{\partial F_a} +
2\pi\mathrm{dt} \frac{\partial f_i(t)}{\partial F_a}
\quad\quad\quad,
}
\end{cases}
\end{equation}	
where $\delta$ is the Kronecker delta, and $\frac{\partial \bm{\theta}(t=0)}{\partial \bm{F}}=\bm{0}$. The derivatives $\frac{\partial \bm{f}(t) }{\partial \bm{F}}$ are then averaged on the last $\tau/\mathrm{dt}$ timesteps to obtain the Jacobian $\frac{\partial \bm{\bar f}}{\partial \bm{F}}$. This algorithm has a complexity in $\mathcal{O}\left(N^3\right)$ but is highly parallelizable.
}

Following standard practices in gradient propagation through time, we implement gradient clipping~\cite{pascanu_difficulty_2013} by applying the $\tanh$ squashing function to the Jacobian. Similarly, the natural frequencies $\bm{F}$ are min-max clipped to stay within $[500;680] \mathrm{MHz}$, which represents the typical tuning range of spin-torque nano-oscillators~\cite{vodenicarevic_nanotechnology-ready_2017}. Unwanted sensitivity to initial conditions is prevented by setting the initial oscillator phases $\bm{\theta_0}$ to uniformly random values in $[0;2\pi)$ for each simulation run. 
The differentiable oscillator network operator can be implemented as a neural layer within popular machine learning libraries for use with automatic differentiation. In this work, we use a Graphics Processing Unit (GPU)-accelerated custom C++ differentiable oscillator network operator for the machine learning library TensorFlow~\cite{tensorflow2015-whitepaper}. The Adam optimizer is used with default parameters and a learning rate of 0.03 in order to minimize the total error $E^\mathrm{tot}$ on a set of training examples.
{\color{black} It should be noted that we assumed a basic Euler numerical integration scheme, but that our approach can be adapted to other explicit integration methods as well.}

\section{Implementation of the Offline learning Strategy}

We {\color{black} first} compare the online learning algorithm to the offline learning algorithm, on the $N_I=2$, $N_T=4$, $k=10\mathrm{MHz}$ reference classifier, by solving a basic binary classification task consisting in synchronizing tuning oscillators $(4,5)$ for a set of positive examples, and desynchronizing them for a set of negative examples. The total number of examples is $M=1156$, of which $86\%$ are negative.

\begin{figure}[ht!]
\includegraphics[]{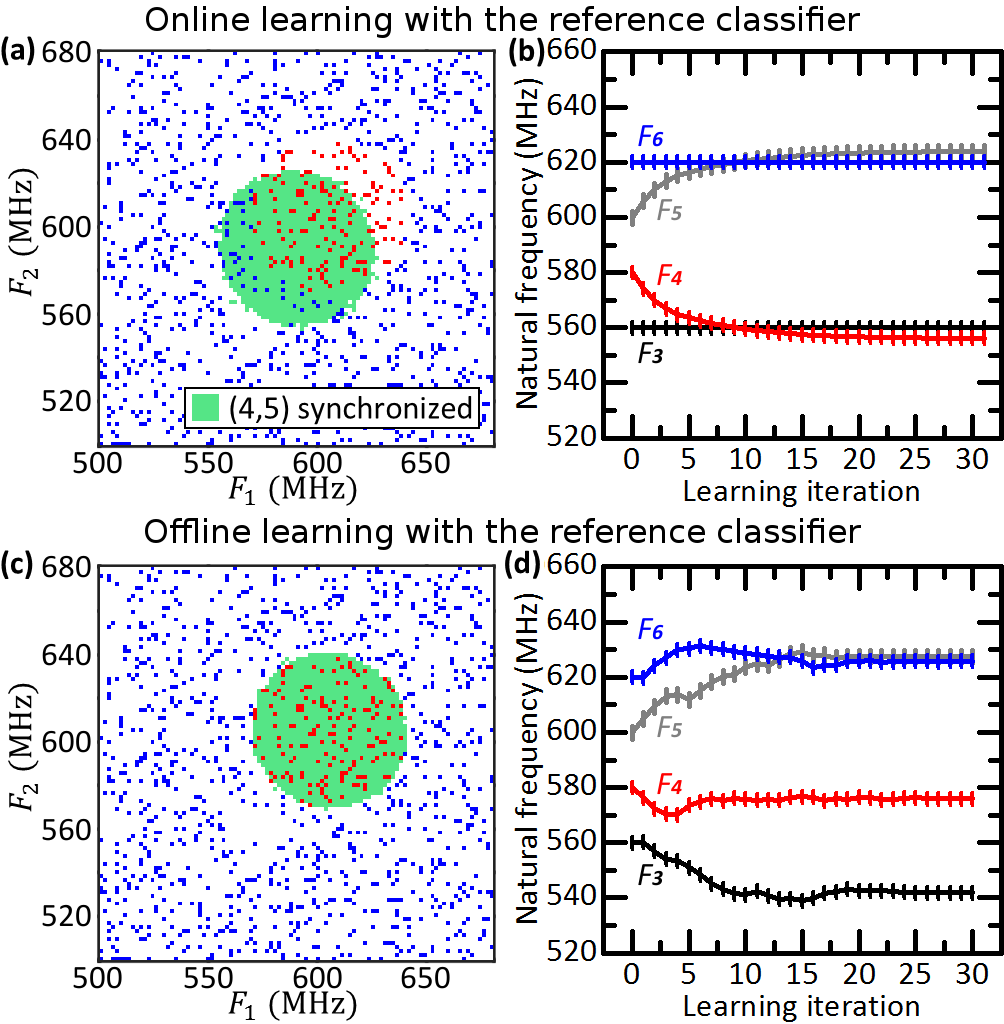}
\caption{Comparing the online and offline learning algorithms on a basic classification task using the reference classification architecture with $N_I=2$ and $N_T=4$. (a) Response map after 20 iterations of online learning. (b) Tunable natural frequencies as functions of the online learning iteration. (c) Response map after 20 iterations of offline learning. (d) Tunable natural frequencies as functions of the offline learning iteration.}
\label{fig:learning}
\end{figure}

Fig.~\ref{fig:learning}(a) shows the response map of the reference classifier after 20 online learning iterations as a function of the input natural frequencies $F_1$ and $F_2$, where oscillators $(4,5)$ are synchronized in the green area, and the red and blue dots represent the positive and negative examples, i.e. the target regions for synchronization and desynchronization.
Fig.~\ref{fig:learning}(b) shows the natural frequencies of the tuning oscillators as functions of the learning iteration.
After 20 iterations, the natural frequencies stabilize, and a classification rate of $91.7\%$ is reached. 
The results show that classification task is not fully solved due to the fact that the only degree of freedom of the online learning algorithm is to push $F_4$ and $F_5$ apart or closer together, thereby adjusting the size of the synchronization region without being able to shift it towards the target region.

In Fig.~\ref{fig:learning}(c) we show the response map of the same classifier after 20 offline learning iterations, and Fig.~\ref{fig:learning}(d) shows the natural frequencies of the tuning oscillators as functions of the learning iteration.
The results show that the offline learning algorithm is able to successfully adjust all four tunable natural frequencies, shift the synchronization region along the first diagonal of the map, and reach an improved classification rate of $99.7\%$.

\begin{figure}[ht!]
\includegraphics[]{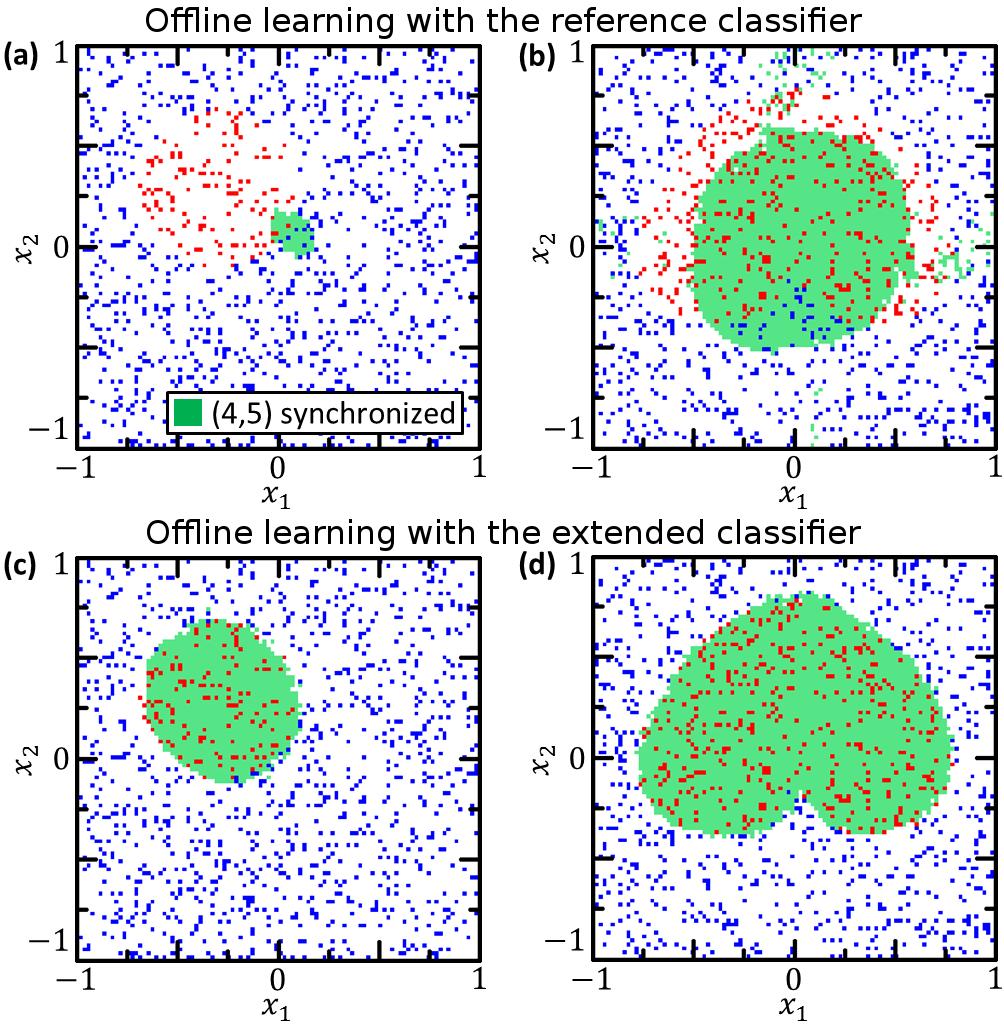}
\caption{Response maps on two different target regions after 100 offline learning iterations for (a,b) the $N_I=2$, $N_T=4$ reference classifier, and (c,d) the $N=6$ extended classifier. {\color{black} $x_1$ and $x_2$ are the components of the bi-dimensional input vector $\bm{x}$.} }
\label{fig:complex_shapes}
\end{figure}

\begin{figure}[ht!]
\includegraphics[]{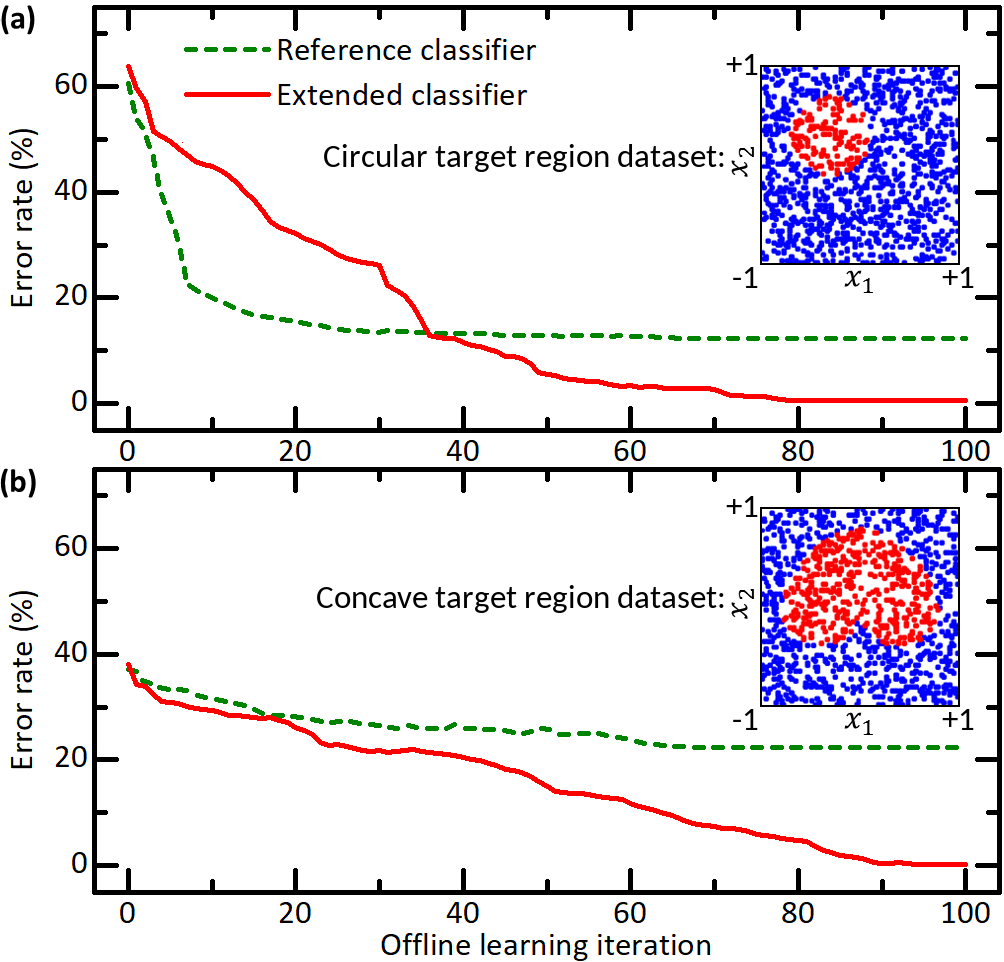}
\caption{\color{black} Error rates (percentage of misclassified examples) as functions of the offline learning iteration, obtained with the $N_I=2$, $N_T=4$ reference classifier and the $N=6$ extended classifier on (a) the off-diagonal circular target region dataset and (b) the convex target region dataset. }
\label{fig:err_rates}
\end{figure}

We then compare the reference classifier ($N_I=2$, $N_T=4$) to the  extended classifier ($N=6$),  with $k=10\mathrm{MHz}$ on more advanced binary classification tasks using the offline learning algorithm.
Figs.~\ref{fig:complex_shapes}(a,b) show the response maps of the reference classifier after 100 offline learning iterations for an off-diagonal circular target region ($M=1156$, $87.5\%$ negatives), and a concave target region ($M=1156$, $59.9\%$ negatives) respectively. Figs.~\ref{fig:complex_shapes}(c,d) show the results obtained in the same conditions with the extended classifier. {\color{black} Figs.~\ref{fig:err_rates}(a,b) show the evolution of the error rates of both classifiers on the two datasets.}
Our results show that the reference classifier is unable to classify non diagonally-symmetric patterns ($87.7\%$ on the off-diagonal circle) due to symmetries, and its low number of tunable parameters limits its capability to classify complex shaped regions ($77.6\%$ on the concave region). The extended classifier, however, successfully learns to represent both the off-diagonal circular ($99.4\%$) and the concave ($99.8\%$) target regions.

These results show that the extended classifier, together with the offline learning algorithm offers superior classification capabilities compared to the reference classifier with online learning.

\section{Benchmarking the extended classifier with offline learning}

We now benchmark the extended classifier ($N=6$) against the  reference classifier ($N_I=4$, $N_T=2$, the two tuning oscillators are used as outputs) using the offline learning algorithm, and a perceptron linear classifier on the standard Iris dataset~\cite{fisher_use_1936}. The binary classification task consists in identifying whether or not a flower belongs to a specific species given four features. We set $k=10\mathrm{MHz}$ and use 30 examples from each of the three species as a 90-item training set, and the 20 remaining examples per species as a 60-item test set.

\begin{table}[ht!]
\centering
\begin{tabular}{|c c c c|} 
\hline
Architecture & \textit{Iris setosa} & \textit{Iris versicolor} & \textit{Iris virginica} \\ 
\hline 
Reference & 67~\% & 87~\% & 67~\% \\ 
\textbf{ Extended}   & \textbf{ 100~\%} & \textbf{ 98~\%} & \textbf{ 100~\%} \\
Perceptron & 100~\% & 73~\% & 97~\% \\
\hline
\end{tabular}
\caption{Iris test set classification rates on the flower species identification task for the three Iris species, using the $N_I=4$, $N_T=2$ reference classifier, the $N=6$ extended classifier, and a perceptron.}
\label{tab:iris}
\end{table}

Table~\ref{tab:iris} shows the test set classification rates on identifying each of the three species after 100 learning iterations for the reference classifier, the extended classifier, and the perceptron.
The results show that the perceptron successfully classifies the linearly separable classes \textit{Setosa} ($100\%$) and \textit{Virginica} ($97\%$) but fails on the non linearly separable \textit{Versicolor} class ($73\%$). On the contrary, the reference classifier fails to classify \textit{Setosa} ($67\%$) and \textit{Virginica} ($67\%$) but performs better on \textit{Versicolor} ($87\%$). {\color{black} This observation highlights the different, and complementary nature of oscillator-based classifiers compared to classical neural networks. Those differences are due to the highly non-monotonic nature of the oscillator network operator and its rich inter-unit interactions, both of which are unusual properties in classical neural network layers. }
Finally, the extended classifier successfully classifies all three species ($100\%$, $98\%$, $100\%$), which shows that its rich and tunable physics provide extra computing power compared to {\color{black}classical} single-layer neural networks.

\begin{figure}[ht!]
\includegraphics[]{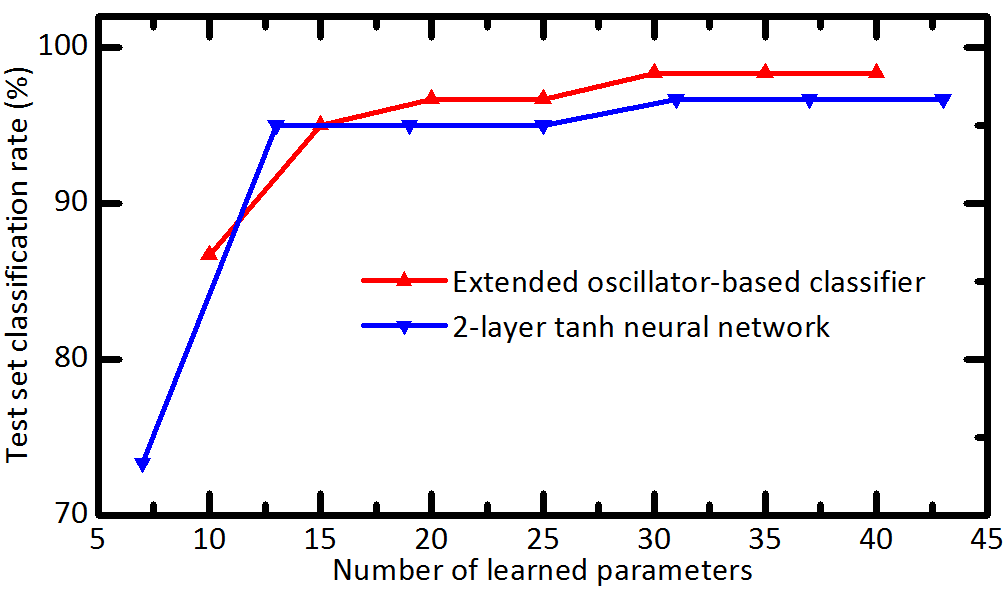}
\caption{\textit{Iris versicolor} identification test set classification rate as a function of the number of learned parameters, for the extended classifier trained offline, and for a two-layer $\tanh$ neural network.}
\label{fig:nparams}
\end{figure}

We then evaluate how well the extended classifier with offline learning scales on the non linearly separable \textit{Iris versicolor} identification task when the number of oscillator increases, and compare it to a two-layer $\tanh$ feed-forward neural network.
Fig.~\ref{fig:nparams} shows the test set classification rates after 100 learning iterations as functions of the number of learned parameters for the extended classifier and the two-layer neural network.
The results show that the extended classifier, using a single weight matrix, 
outperforms the two-layer neural network,  which uses two weight matrices, at equivalent number of learned parameters for 15 learned parameters and above.

\begin{figure}[ht!]
\includegraphics[]{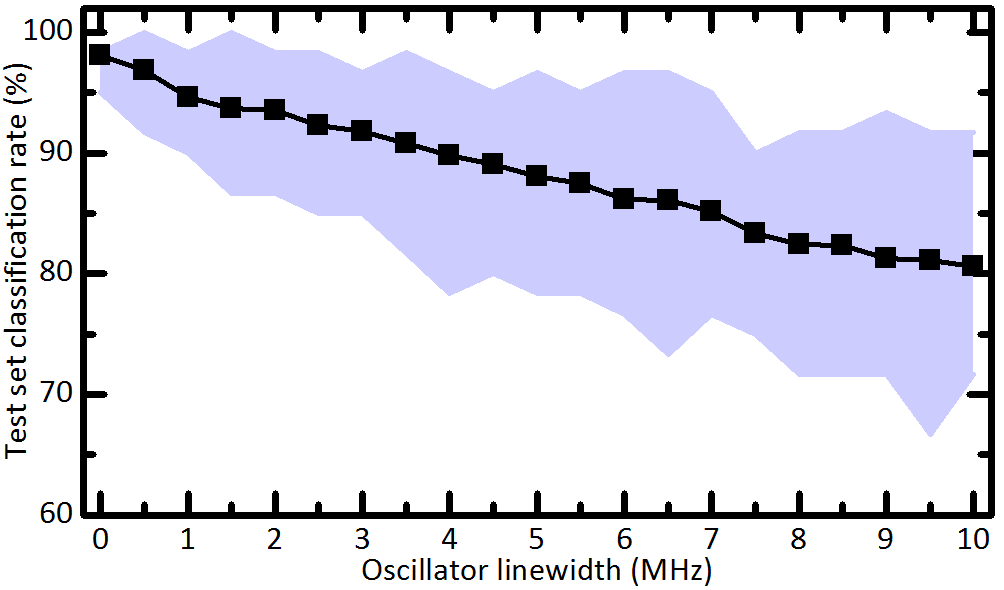}
\caption{Average \textit{Iris versicolor} identification test set classification rates obtained on 100 trials of a noisy extended classifier trained using the offline learning algorithm, as a function of the oscillator linewidth. The span between the worst and best cases encountered during the trials is blue-filled.}
\label{fig:noise}
\end{figure}

Real nano-oscillator systems are subject to phase noise, which is not included during offline learning in order to keep meaningful gradients. It is therefore important to evaluate the effects of phase noise when using a weight matrix trained offline in a noisy physical system.
We choose to train the $N=6$ extended classifier on the \textit{Iris versicolor} identification task with the offline learning algorithm for 100 learning iterations. The obtained weight matrix is then transferred into a simulated noisy Kuramoto model of the network, and the test set classification rates are evaluated for different noise levels expressed as the linewidth of an isolated oscillator. {\color{black} The noisy version of the phase update equations~(\ref{eq:euler}) is:
\begin{equation}\label{eq:eulernoise}
\begin{cases}
\displaystyle{f_i(t)= F_i + k\sum_{i=1}^N{ \sin{\left(\theta_j(t) - \theta_i(t)\right)} }}\\[1.2em]
\displaystyle{\theta_i(t+\mathrm{dt})= \theta_i(t) + 2\pi \mathrm{dt} f_i(t)  + \sqrt{ 2\pi \mathrm{dt} \Delta }\,\,\,\mathcal{N}}\\[0.5em]
\end{cases} \,,
\end{equation}
where $\Delta$ is the linewidth of an isolated oscillator, and $\mathcal{N}$ is a pseudo-random Gaussian distribution with mean $0$ and variance $1$.
}
For each noise level, the system is simulated 100 times with different random initial phases and different noise random seeds.
Fig.~\ref{fig:noise} shows the average test set classification rate obtained with the noisy system on 100 trials, as a function of the oscillator linewidth. The blue-filled area represents the span between the worst and best test set classification rates encountered during the trials.
The results show that noise decreases the classification rates. However, for noise levels typical of spin-torque nano-oscillators ($\Delta \approx 1 \mathrm{MHz}$~\cite{vodenicarevic_nanotechnology-ready_2017}), the average classification rate is $94.6\%$ with a worst case at $90\%$, showing that the offline learning approach is robust to realistic levels of phase noise.

\section{Discussion}

This work defines the first full-featured supervised offline learning algorithm for oscillator-based classifiers, as well as an improved classification architecture compatible with current nanotechnologies. Comparison with the reference approach of \cite{vassilieva_learning_2011} showed the power of the approach in terms of classification accuracy, while comparisons with standard neural networks showed that  our approach was economical in terms of adjustable parameters.

The main benefit of our approach is its compatibility with both the constrains of nanotechnology (no tuning of the coupling \textit{between} the oscillators) and with the standard techniques of machine learning. The neural network can be trained offline, using the powerful frameworks of software neural networks, such as ~\cite{tensorflow2015-whitepaper} used here.
By contrast, the main limitation of our approach also comes from the  offline nature of learning. There can be mismatch between the model used for training, and the reality of the physical system in which the trained parameters are transferred.
The first mismatch mitigation approach is of course to choose a realistic device model. This work relies on the Kuramoto model which is used to describe a wide range of oscillators~\cite{acebron_kuramoto_2005}, but any other smooth model~\cite{locatelli_vortex-based_2015} can be used instead to produce trained weight matrices that are better fit to the target oscillator system.
Mismatch effects can also be mitigated by adding random variability on the model properties at each learning step in order to avoid over-fitting and produce robust solutions that transfer to variable physical devices.
The impact of mismatch could also be further reduced by performing extra online learning steps to perfect the general solution provided by the offline learning algorithm. 

{\color{black} It should also be noted that the training process is relatively heavy, and slower than for more conventional neural networks, due to the added complexity of modeling the dynamics of the nanooscillators. In terms of application, our approach therefore targets inference hardware, where the model is trained on a server and then programmed on highly efficient specialized circuits provided to users.
}

Furthermore, this study focuses on binary classification tasks, but multi-class classification can be readily achieved either by training one binary classifier per class, or by using multiple pairs of oscillators as outputs.
Oscillator networks can also be stacked into multi-layer architectures and trained transparently with the offline learning algorithm thanks to automatic differentiation.

Finally, even though we only used the average stable state of the oscillator network, the offline learning algorithm can be applied to temporal signals as well, by inputting a different $\bm{F}$ at every time-step, and reading a sliding time window average of $\bm{f}(t)$. A promising research direction would be to exploit the temporal and transient behaviors of the oscillator network to classify or generate sequences and signals in time.

In conclusion, we showed that neural networks based on nano-oscillators, despite their nature that differs strongly from standard neural networks, can be trained using similar frameworks and methodology, to achieve good classification performance. This work suggests the value of bringer closer works performed in physics and in machine learning for the implementation of compact and energy efficient classification systems.

\begin{acknowledgments}
This work is supported by a public grant overseen by the French National Research Agency (ANR) as part of the Investissements d'Avenir program (Labex NanoSaclay, reference: ANR-10-LABX-0035), by the ANR MEMOS grant (reference: ANR-14-CE26-0021) and by the French Minist\`ere de ́l'\'ecologie, du d\'eveloppement durable et de l'\'energie.
The authors thank P.~Talatchian, M.~Romera, F.~A.~Araujo, A.~Mizrahi, A.~F.~Vincent, T.~Hirtzlin, M.~Ernoult, and C.~H.~Bennett for fruitful discussions.
\end{acknowledgments}



%

\end{document}